\documentstyle[aps,twocolumn,floats]{revtex}

\def\e{\tilde{\eta}}
\def\del{\partial}
\def\be{\begin{equation}}
\def\ee{\end{equation}}
\def\bea{\begin{eqnarray}}
\def\eea{\end{eqnarray}}

\def\II{I\hspace{-.1em}I\hspace{.1em}}

\def\be{\begin{equation}}
\def\ee{\end{equation}}
\def\bea{\begin{eqnarray}}
\def\eea{\end{eqnarray}}

\begin{document}

\tighten
\draft
\twocolumn[\hsize\textwidth\columnwidth\hsize\csname 
@twocolumnfalse\endcsname

\title{ Noncommutative Spacetime,
Stringy Spacetime Uncertainty Principle, \\
and Density Fluctuations }

\author{Robert Brandenberger}

\address{TH Division, CERN, CH-1211 Geneva 23, Switzerland\\
and\\
Department of Physics, Brown University, Providence, RI 02912, USA\\
E-mail: rhb@het.brown.edu}

\author{Pei-Ming Ho}

\address{
Department of Physics, National Taiwan University, Taipei 106, Taiwan, R.O.C.\\
E-mail: pmho@phys.ntu.edu.tw}

%

\maketitle

\begin{abstract}

We propose a variation of
spacetime noncommutative field theory
to realize the stringy spacetime uncertainty relation
without breaking any of the global symmetries
of the homogeneous isotropic universe.
We study the spectrum of metric perturbations in this model
for a wide class of accelerating background cosmologies.
Spacetime noncommutativity leads to a coupling between
the fluctuation modes and the background cosmology
which is nonlocal in time. For each mode, there is
a critical time at which the spacetime uncertainty
relation is saturated. This is the time when the mode
is generated. These effects lead to a
spectrum of fluctuations whose spectral
index is different from what is obtained for
commutative spacetime in the infrared region,
but is unchanged in the ultraviolet region.
In the special case of an exponentially expanding
background, we find a scale-invariant spectrum.
but with a different magnitude than in the context
of commutative spacetime if
the Hubble constant is above the string scale.

\end{abstract}

\vspace*{1cm}
]

\section{Introduction}

As a candidate for the theory of everything,
string theory should tell us everything about the universe.
One of the most important questions
is how a successful theory of cosmology
can be derived from it.
Cosmology is becoming the major testing ground
for string theory,
since it tests the physics at energies much higher
than can be reached in any collider on earth.
On the other hand, cosmology also appears to require the
input from string theory, since
cosmological theories based on
classical gravity and the standard model of particle physics
are expected to break down at very high energies (see e.g.
\cite{RHBrev} for a discussion of some of the conceptual
problems of inflationary cosmology).
There is a growing symbiotic relationship between string theory and 
cosmology (see e.g. \cite{RHBAPCTP}).

In this paper we are concerned with
what string theory can say about cosmology.
Obviously, we are very far from being able
to derive cosmology directly from string theory.
Instead of proposing a new scenario starting from
a very special string theory configuration,
we would like to focus on 
a universal property of string theory,
and study its implication for cosmology.
The universal property which we wish to focus on
is the stringy spacetime uncertainty relation (SSUR),
which states that the uncertainties in
the physical time and space coordinates 
$\Delta t_p = \Delta t$ and $\Delta x_p$ satisfy
\be \label{ssur}
\Delta t_p \Delta x_p \geq l_s^2,
\ee
where $l_s$ is the string length scale.
Yoneya suggested \cite{Yoneya} that the SSUR
is a universal property for strings as well as D-branes.
This is in contrast with
the other stringy uncertainty relation \cite{OSUR}
\be \label{ur}
\Delta x_p \Delta p\geq 1+l_s^2\Delta p^2,
\ee
which implies a minimal length scale
\be \label{dx}
\Delta x_p \geq l_s
\ee
but does not hold for D-branes.
In an earlier paper \cite{GHR},
it was argued that the SSUR
can be used to solve the flatness and
the horizon problems in cosmology
without the need to invoke cosmic inflation.

In this paper,
we study the effect of the SSUR on
metric perturbations in the early universe,
which are the origin of the observed large-scale
structure and of cosmic microwave background anisotropies.
The reason one expects signatures of
Planck (string) scale physics in the spectrum of
density fluctuations on cosmological scales today
is the following \cite{RHBrev,MB}:
since the fluctuations on large scales are small today,
and since gravity is a purely attractive force,
the fluctuations had to be extremely small in the
early Universe, thus justifying a linearized analysis.
According to such an analysis,
the individual Fourier modes of the fluctuation field
evolve independently.
In the context of an expanding background cosmology,
there was thus a time when the physical wavelengths
of modes which are probed in current cosmological
experiments was smaller than the Planck (or string) scale,
and hence one cannot neglect the effects of
Planck (string) scale physics
\footnote{In this context of inflationary cosmology,
this non-robustness of the predictions of inflation
to possible effects of Planck (string) scale physics
is known as the {\it trans-Planckian problem}.}.

As a first step, we consider a simple model in
which matter is dominated by a single scalar field.
As is done in inflationary cosmology,
we quantize the joint linear metric and matter fluctuations
about a classical homogeneous and isotropic background cosmology.
With these assumptions the cosmological perturbations
are automatically Gaussian.

In order to carry out the calculation explicitly,
we need a specific model to realize the SSUR.
Motivated by the recent developments
on noncommtuative geometry in string theory,
we generalize the field theory
on 1+1 dimensional noncommutative space,
where the uncertainty relation is
a direct result of the spacetime noncommutativity,
to (3+1) dimensions.

We identify two crucial effects which lead to a difference
between the evolution of fluctuations in commutative and
noncommutative spacetimes. The first is a coupling between
the fluctuation mode and the background which is nonlocal
in time, and the second is the appearance of a critical
time for each mode at which the SSUR is saturated, and which
is taken to be the time when the mode is generated.
\footnote{Note that this implies that the Hilbert space
of the perturbative quantum theory of gravity becomes
time-dependent. This appears to be a natural consequence
of any attempt to quantize gravity in the context of
cosmology, as discussed in \cite{Jacobson:1999zk} and
in references therein.}

These two effects lead to a change in the spectral index of
the fluctuations in the infrared region for all accelerating
background cosmologies except for the exponentially expanding
case. Instead of a red spectrum (as is obtained in
commutative spacetime) we obtain a blue spectrum. In the
case of an exponentially expanding background, a scale-invariant
spectrum results.

\section{Preliminaries}

In string theory,
the best known example of noncommutative field theory
is the low energy effective theory of D-branes
in the background of a $B$ field \cite{SW}.
It has also been proposed that
R-R background fields may lead to
a low energy effective theory of gravity
best described as Einstein gravity
living on a certain noncommutative spacetime \cite{JR}.
Motivated by recent developments
of noncommutative geometry in string theory,
models for cosmological fluctuations on noncommutative spacetime
have been proposed by several groups \cite{CGS,ABM}.
\footnote{
Models based on (\ref{ur}) are also studied \cite{Kempf,EGKS}.
}
However, in these proposals,
some isometries of the FRW metric
are broken by the noncommutativity.
In this paper,
we will not give a specific commutation relation
for spacetime.
Rather, we will imagine that the SSUR \cite{Yoneya}
is realized by a deformation of the commutative field theory
in a way which preserves all of the global symmetries
of the classical background.
One can imagine that,
although their background expectation values vanish
and thus there is no well-defined spacetime noncommutativity,
quantum fluctuations of the $B$ field (for D-branes)
or certain R-R fields (for spacetime)
would still result in spacetime uncertainty \cite{CHK}.
Keeping this in mind,
we will first consider field theories
on noncommutative space,
which may give us a hint about the result of
``averaging over'' (in a path integral)
the effective action for different background fields.

The formulation of general relativity
on noncommutative spacetime
has been extensively studied \cite{Connes,CFF,Ho}.
The notions of metric, distance, etc. can all be defined.
It is therefore natural for us to assume that
the notion of metric still makes sense
as long as it is consistent with SSUR.
Assuming a homogeneous isotropic background,
we take the Friedmann-Robertson-Walker (FRW) metric
\bea \label{FRW}
ds^2 =
dt^2-a^2(t)\left(\frac{dr^2}{1-K r^2}-r^2 d\Omega^2\right).
\eea
For simplicity, we will focus on the case
of a spatially flat universe ($K=0$).
For later use, we introduce another time coordinate $\tau$
(not the usual conformal time) such that the metric is
\be
ds^2 = dt^2 - a^2(t) dx^2
= a^{-2}(\tau)d\tau^2 - a^2(\tau) dx^2.
\ee

Since the scale factor has no spatial dependence,
the SSUR imposes no restriction on
the Hubble constant $H\equiv \dot{a}/a$,
or any other scale associated with the scale factor,
such as $\dot{H}^{1/2}$.
On the other hand,
the constraint (\ref{dx})
implies that $H < M_s$, and $\dot{H} < M_s^2$, etc.,
where $M_s = l_s^{-1}$ is the string energy scale.
As we will see below,
this difference is crucial for us. 

In view of the connection between
noncommutative field theory and SSUR described above,
the fact that SSUR does not impose a restriction
on quantities with only time dependence can
be understood via noncommutative field theory.
Obviously, if we only look for solutions
with a single variable,
noncommutative field theories are equivalent to
commutative field theories.

In the case of a classical spacetime,
for a given equation of state of matter $p=w\rho$
with constant $w$,
the solution of the (classical)
Einstein equations for the scale factor is
\be \label{a}
a(t) = a_0 t^n = \alpha_0 \tau^{n \over n+1},
\ee
where
\be
n = { 2 \over 3(1+w) }
\ee
and $\alpha_0 = \left( (n+1)^n a_0 \right)^{ 1\over n+1}$.
For an expanding universe, $n > 0$.
For a matter dominated universe,
$w = 0$ and $n = 2/3$;
for a radiation dominated universe,
$w = 1/3$ and $n = 1/2$.
The universe accelerates if $n > 1$.
If the cosmological constant dominates, then
$w = -1$, $a(t) = a_0 \exp(Ht)$ ($n = \infty$)
and we have exponential inflation \cite{Guth,Linde}.

Since we do not have the exact gravitational theory
with all the stringy corrections to Einstein gravity,
we will consider a generic expanding scale factor $a$.
As we review in the Appendix,
for a given scale factor $a$,
the metric perturbations
for gravity coupled to a single scalar field
obey an equation of motion of the form
\be
\mu^{''}_k+\left(k^2-\frac{z^{''}_k}{z_k}\right)\mu_k=0,
\ee
where the primes mean derivatives with respect to
a certain time coordinate $\e$ to be defined below.
(In the commutative case, $\e = \eta$
is the conformal time defined by $dt = a d\eta$.)
This equation applies both to gravitational waves
(in which case $\mu = a h$, $h$ being the amplitude of the
gravitational wave, and - for commutative spacetime - $z_k = a$)
and to scalar metric fluctuations,
the fluctuations which couple to matter
(in which case $a^{-1} \mu$ is the scalar field
fluctuation in the uniform spatial curvature gauge and
- in commutative spacetime - $z_k = z$ 
is a function describing the background model 
and is given explicitly by (\ref{zdef})).
For backgrounds in which the equation of state
does not change in time (the ones considered here),
the function $z$ for scalar metric fluctuations
is proportional to $a$,
and hence the equations for scalar metric fluctuations
and gravitational waves coincide.

A quantity used frequently to compare theory
with observations is the power spectrum,
which measures the strength of the 
fluctuations on the scale $k$.
The power spectrum of metric perturbations is given by
\be \label{Pk1}
P_k = {{k^3} \over {2 \pi^2}}\frac{|\mu_k|^2}{z_k^2} \, .
\ee
(To allow comparison with what is done for
a non-commutative spacetime, we have allowed $z$ to depend on $k$)
If the background is expanding, then as explained in the Appendix
(with the normalization factor chosen to apply to gravitational
waves),
\be \label{Pk2}
P(k) \simeq k^2 z^{-2}_k(\e_k) M_P^{-2}
\ee
where $M_P$ denotes the Planck mass, and $\e_k$ is the earliest time when
\be \label{etak}
\frac{z^{''}_k}{z_k} > k^2 .
\ee

In the usual expanding cosmological models
described by the ansatz (\ref{a}), then 
for a commutative spacetime,
we have $z_k = a$. Hence, for $n > 1$, the time $\e_k$ is the time when the
scale $k$ exits the Hubble radius,
and  we have $\eta_k \propto k^{-1}$,
and thus $z_k(\eta_k)\propto k^{n \over 1-n}$.
Therefore, the spectrum of gravitational waves
and of scalar metric fluctuations obeys
\be \label{oldPk}
P_k\simeq c k^{2 \over 1-n},
\ee
where $c = \left( (2 n^2 - n)^n a_0^2 \right)^{1 \over n-1}$.
For the spectrum to be exactly scale invariant,
we need $n = \infty$, that is, exponential inflation,
for which the spectrum is
\be \label{infl}
P_k\simeq\frac{H^2}{M_P^2}.
\ee

\section{Toy Model on Noncommutative Spacetime}

First we consider the case of 1+1 dimensions
where the SSUR can be easily realized by
the $\ast$-product,
so that we can get some hint about how
the SSUR affects the action of a scalar field.

The SSUR
\be
\Delta t \Delta x_p = \Delta\tau \Delta x\geq l_s^2 \, .
\ee
can be realized by the algebra of noncommutative spacetime
\be \label{taux}
 [\tau, x]_{\ast}=i l_s^2 \, .
\ee
This $\ast$-product can be explicitly defined as
\bea
& & (f\ast g)(x,\eta) = \\
& &\left.
e^{-\frac{i}{2}l_s^2 (\del_{x}\del_{\tau'}-
\del_{\tau}\del_{y})}f(x,\tau)g(y,\tau') \right|_{y=x, \tau'=\tau}.
\nonumber
\eea

The fact that the algebra (\ref{taux})
is time-independent makes $\tau$ a better coordinate
than the conformal time $\eta$ defined by $dt = ad\eta$,
which is frequently used when calculating metric perturbations.
Naively, the SSUR can be written as
\be \label{SSUR1}
\Delta\eta\Delta x \geq \frac{l_s^2}{a^2(\eta)}
\ee
in terms of $\eta$.
However, this relation is not well defined
when $\Delta\eta$ is large,
because the argument $\eta$ for the scale factor
on the right hand side changes over the time interval $\Delta \eta$,
and it is thus not clear what to use for $a(\eta)$ in (\ref{SSUR1}).

A comment on unitarity is in need at this point.
Since spacetime noncommutativity introduces
higher derivatives of time in the Lagrangian of a field theory,
a naive treatment will often result in loss of unitarity.
(But there are also examples \cite{Chu}
for which the unitarity is not broken
by spacetime noncommutativity.)
However, in our case, the fundamental theory is string theory,
and we are looking at its effective field theory
(with focus on the stringy effect described by the SSUR).
It is very common for effective theories to
have higher derivatives terms,
and a proper treatment \cite{EW} should preserve
unitarity order by order perturbatively.
Furthermore, the field theory we will consider below
is essentially a free field theory,
and so we will not have to worry
about unitarity in this paper.

As discussed in the Appendix,
both scalar metric fluctuations and gravitational waves
are described by free scalar field actions
on the classical expanding background
(in two spacetime dimensions there are
obviously no gravitational waves).
We will now propose a noncommutative field theory
which generalizes the action for cosmological perturbations
to the case of noncommutative space-time.
We will start from the actions expressed in terms of
the observable fields $h$ and ${\cal R}$
in the case of gravitational waves and scalar metric fluctuations,
respectively (see Appendix).
In the case of scalar metric fluctuations
the variable ${\cal R}$ is the fluctuation of
the spatial curvature in the comoving gauge
in which $\delta \varphi = 0$.
As can be deduced from (\ref{gravac}) and (\ref{scaleac}), respectively,
these are free field actions except that
the expansion of the background cosmology has not been factored out 
(i.e. the equations of motion for these variables
contain the Hubble damping terms).
These actions thus contain a nontrivial measure factor
$a^{d-1}$ or $z^{d-1}$, in the cases of gravitational waves
and scalar metric fluctuations, respectively,
where $d$ is the number of spatial dimensions.
We will extract a factor of $a^{-2}$ from the
measure and insert it into the operator appearing in
the Lagrangian to account for the fact that
the spatial gradient appearing in the operator
should reduce to the usual operator in Minkowski space-time
when expressed in terms of physical distances.
We propose to make the transition to noncommutative spacetime
obeying the SSUR by taking the operator appearing in
the action and replacing all multiplications by $\ast$-products.

Based on the above considerations, we take 
the free field action for a real scalar field in 1+1 dimensions
\be
\tilde{S}=\int d\tau dx\frac{1}{2}
\left(
\del_{\tau}\phi^{\dagger}\ast a^{2}\ast\del_{\tau}\phi
-(\del_x\phi)^{\dagger}\ast a^{-2}\ast(\del_x\phi)
\right).
\ee
In terms of the Fourier transform of $\phi$,
\be \label{Four}
\phi(\tau, x) = V^{1/2} \int_{k<k_0(\tau)} \frac{dk}{2\pi}
\phi_k(\tau) e^{ikx},
\ee
(where $V$ is the total spatial coordinate volume)
the action is
\be \label{S1}
\tilde{S} \simeq V\int_{k<k_0} d\tau dk \frac{1}{2}
\left(\beta^+_k\del_{\tau}\phi_{-k}\del_{\tau}\phi_k
-k^2\beta^-_k\phi_{-k}\phi_k\right),
\ee
where
\be \label{beta0}
\beta^{\pm}_k(\tau)=
\frac{1}{2}\left(
a^{\pm 2}(\tau-l_s^2 k)+a^{\pm 2}(\tau+l_s^2 k)
\right).
\ee
The $\ast$-product in the action takes care of
the SSUR for the interaction
between the background metric and the scalar field.
In order to realize the SSUR for the scalar field by itself,
we have imposed an upper bound on the comoving momentum
$k$ at $k_0$ in (\ref{Four}).
The reason is as follows.
In order for a fluctuation mode with wave number $k$ to exist,
the SSUR must be satisfied.
According to (\ref{S1}),
the energy defined with respect to $\tau$ for a mode $k$ is
\be
\label{Ek}
E_k = k a^{-2}_{eff},
\ee
where
\be \label{aeff}
a^{2}_{eff} = \left({\beta^+_k \over \beta^-_k}\right)^{1/2}.
\ee
Using $\Delta x \sim 1/k$, $\Delta \tau \sim 1/E_k$ 
together with the SSUR, we find
\be
\left(\frac{a_{eff}(\tau)}{k}\right)^2 \sim
\Delta x_p \Delta t \geq l_s^2
\ee
and thus we have an upper bound on the wave number
\be \label{k0}
k \leq k_0(\tau) \equiv \frac{a_{eff}(\tau)}{l_s}.
\ee
One should also check whether the background metric
satisfies the SSUR by itself.
Yet as we mentioned above,
the SSUR imposes no contraint on the background
since it is homogeneous.

Due to the SSUR,
the mode $\phi_k$ interacts with the background with
an uncertainty of $l_s^2 k$ in $\tau$.
In general, if the uncertainty relation is not realized exactly
by the commutation relation (\ref{taux}),
we expect that $\beta^{\pm}_k$ will be
replaced by functions of the form
\be \label{beta}
\beta^{\pm}_k(\tau)=\int d\xi f(\xi) a^{-2}(\tau-\xi k),
\ee
for some even function $f(\xi)$ peaked at the origin
with a characteristic width $l_s^2$.
By Taylor expansion, we have
\be \label{beta1}
\beta^{\pm}_k(\tau)=
\left( 1 + C^{\pm}_1 {H p \over M_s^2}
+ C^{\pm}_2 {(\dot{H} + H^2) p^2 \over M_s^4} + \cdots \right)
a^{\pm 2}(\tau),
\ee
where $C^{\pm}$ are constants of order $1$,
and $p=k/a$ is the physical momentum.
This is a special case of the most general correction
due to new physics at an energy scale $M_s$,
which can be expanded in powers of $(H/M_s)^2$
and $(p/M_s)^2$ independently \cite{KKLS}.

The action (\ref{S1}) reduces to the action
for metric fluctuations (\ref{gravac})
on classical spacetime
when $l_s\rightarrow 0$.

To calculate the power spectrum,
it is convenient to rewrite the action in the form
\be \label{S2}
\tilde{S} \simeq V\int_{k<k_0} d\e dk \frac{1}{2} y_k^2(\e)
\left(\phi^{'}_{-k}\phi^{'}_k-k^2\phi_{-k}\phi_k\right),
\ee
where the new time coordinate $\e$
is defined by
\be
\frac{d\e}{d\tau} = \left(\frac{\beta^-_k}{\beta^+_k}\right)^{1/2}
= a^{-2}_{eff},
\ee
and
\be
\label{yk}
y_k = (\beta^-_k \beta^+_k)^{1/4}.
\ee
The primes mean derivatives with respect to $\e$.

\section{The Model}

The previous section motivates a model to
incorporate the SSUR for any spacetime dimension:
\be \label{action}
S = V \int_{k<k_0}
d\e d^d k \frac{1}{2}z^{d-1}_k(\e)\left(
\phi^{'}_{-k}\phi^{'}_k-k^2\phi_{-k}\phi_k\right),
\ee
where $z_k$ is some smeared version of $z$ or $a$
over a range of time of characteristic scale
$\Delta\tau = l_s^2 k$.
As $\tau$ increases, the effect of the shift
$\Delta\tau$ for a given mode becomes less important.
The time coordiante $\e$ is related to $\tau$
by $d\e = \tilde{z}^{-2}_k d\tau$,
where $\tilde{z}_k$ is another smeared version of $a$,
As an example, suppose that the only difference between
the $d+1$ dimensional action and the $1+1$ dimensional one (\ref{S2})
is the measure $z^{d-1}$ for the additional $(d-1)$ dimensions,
then we have
\be \label{zkdef}
z^{d-1}_k(\e) = z^{d-1} y_k^2(\e), \quad
\tilde{z}_k(\e) = a_{eff}(\e)
\ee
with $y_k$ given by (\ref{yk}), and $a_{eff}$ by (\ref{aeff}).
In the case of gravitational waves,
the function $z_k$ is denoted $a_k$,
with $a_k$ constructed from the scale factor $a$ in
the same way as $z_k$ is obtained from $z$, say, in (\ref{zkdef}).

This deformation has the advantage of preserving
both spatial translational and rotational symmetry
of the (flat) FRW metric,
in constrast with constructions based on
the commutation relations
\be
[x^{\mu}, x^{\nu}]=i\theta^{\mu\nu} .
\ee

We can now turn to the calculation of the spectrum of
cosmological fluctuations in various
expanding background cosmologies,
including inflationary backgrounds.
We emphasize here that
except the flat FRW metric,
so far we have not assumed anything about
the stringy correction to the Einstein gravity.
A key role in which the SSUR enters in
the analysis of fluctuations
is the existence of a characteristic time
$\e^0_k$ for each mode $k$
which is the time when the SSUR is saturated
\be \label{eta0k}
k=k_0(\e^0_k).
\ee
According to (\ref{k0}),
the mode k can not exist before $\e_k^0$.

A major issue is in which state the fluctuations are generated.
On scales which are smaller than the Hubble radius
at the time of formation, the distinguished choice
is the local vacuum state (the state which
appears empty of particles in the comoving frame
at the time of formation).
But it is unclear when should be the time
to impose this initial condition.
For modes which are generated when the wavelength is
greater than the Hubble length,
the choice is usually even less clear.
However, in our case, the choice is obvious for all modes.
At $\e < \e^0_k$, the fluctuation mode $k$ does not exist.
By continuity of $N_k$, the operator representing the
number of quanta of the k'th mode measured with respect to
the adiabatic vacuum (see e.g. \cite{BDbook} for a textbook
discussion of these concepts), then when the
mode first becomes physical at $\e = \e^0_k$,
it must be in the vacuum state.
Our choice is thus
to consider the amplitude of the growing mode
at the time of formation to be the same as the function
would have in the vacuum state in the absence of
cosmological expansion.

As will be derived below, the fluctuations have a
different spectral index in two wavelength regions which
we call UV and IR, respectively. The UV modes are generated within the 
Hubble radius, and it turns out that for these modes 
the approximation (\ref{app1}) holds, 
The IR modes are generated outside the Hubble radius,
where the effect of the SSUR is important, and the
approximation (\ref{app1}) breaks down.

We first consider the {\bf UV region}, namely 
values of $k$ for which $\Delta \tau$ is small
in the sense that all smeared versions of $a$
can be approximated by $a$.
As in (\ref{beta1}),
the smeared scale factors $z_k$, $\tilde{z}_k$ are expansions
of $( H p / M_s^2 )^2$ (and $\dot{H} p^2 / M_s^4 )$ etc.)
The approximation
\be \label{app1}
z_k \simeq \tilde{z}_k \simeq a
\ee
is valid when
\be \label{bound1}
H p \ll M_s^2.
\ee
As $a$ increases with time,
this approximation gets better and better.

For this bound to be true at all times for a mode $k$,
we only need to make sure that it is satisfied
when the mode first appeared at $\e^0_k$.
Since $\Delta x_p = \Delta t_p \simeq 1/p$,
the saturation of the SSUR at $\e^0_k$
implies that
\be
p =  M_s
\ee
at $\e^0_k$.
Hence,
(\ref{bound1}) is equivalent to
\be \label{bound2}
H(\e^0_k) \ll M_s.
\ee
For the ansatz (\ref{a}), as an example,
the approximation (\ref{app1}) is good if
\be \label{b1}
k \gg A \equiv \alpha_0^{n+1} l_s^{n-1}
\ee

From (\ref{Pk2}),
the perturbation spectrum is determined by
the earliest time when (\ref{etak}) is satisfied.
In the absence of the cutoff (\ref{k0}),
this is the time when
\be \label{etak1}
\frac{z^{''}_k}{z_k} \simeq k^2 .
\ee
Assuming (\ref{app1}),
this condition can be rewritten as
\be \label{ek1}
\dot{H} + 2 H^2 \simeq p^2.
\ee
For $n > 1$ (the accelerating case),
this occurs roughly when the size of
the fluctuation crosses the Hubble radius, i.e.,
\be
H(\e_k) \simeq p.
\ee
From (\ref{bound2}), we see that $\e^0_k \leq \e_k$.
This means that the fluctuations are generated inside
of the Hubble radius, and this was the initial
definition of the UV region. Thus, it is presicely
in the UV region 
that the approximation (\ref{app1}) is valid. 

Since the UR modes are generated on scales inside the Hubble
radius in their local vacuum state, and since the evolution
of the modes after that is no different than in the case
of a commutative spacetime,
it follows immediately that the spectrum for the UV modes (\ref{b1})
is the same as the classical case (\ref{oldPk}).
The uncertainty relation has no significant effect
for these modes,
in agreement with the general consideration of \cite{KKLS}.

Now we would like to study the {\bf IR modes} $k \ll A$
which are generated outside the Hubble radius,
and for which the effects of the SSUR are important.
Although our discussion about the IR modes
will be more speculative
because the Hubble expansion rate is
above the string scale at the time of IR mode formation,
the following provides an interesting example showing
how new physics at the string scale can
have significant effect on the spectrum of perturbations,
evading the pessimistic conclusion of \cite{KKLS}.
\footnote{
After completion of this manuscript two papers
appeared which reach similar conlusions concerning the
potential observability of trans-Planckian physics in
the spectrum of fluctuations \cite{new1,new2}.
}
As we will see, our description of the SSUR
results in a spectrum which except
in the case of exponential inflation has a different
spectral index than what would be obtained in commutative
space-time. 

As an example, we take
$z_k$ and $\tilde{z}_k$ given by (\ref{zkdef}) for $z = a$.
The first step in the analysis is to obtain the expression
for $\tau^0_k$. Starting point is (\ref{k0}) which defines
the initial time. Inserting the expression (\ref{aeff}) for
$a_{\rm{eff}}^2$ and the formulas (\ref{beta0}) for $\beta_k^{\pm}$,
and then using the equation (\ref{a}) for the scale factor, we
obtain
\be \label{inittime}
\tau^0_k = \left( \left( k l_s \over \alpha_0 \right)^{2(n+1)/n}
+k^2 l_s^4 \right)^{1/2}.
\ee
For the IR  modes $k$ which we are focusing on, i.e. $k \ll A$,
the second term dominates over the first term.
Thus, we have $\beta^+_k \simeq {1 \over 2}a^2(\tau+l_s^2 k)$
and $\beta^-_k \simeq {1 \over 2}a^{-2}(\tau-l_s^2 k)$
(in the case of an expanding universe).

To find the power spectrum for these modes, we start from (\ref{Pk2})
with the time ${\tilde{\eta}_k}$ replaced by the time ${\tilde{\eta}_k^0}$
when the modes are generated. We then use (\ref{zkdef}) to
replace $z_k^2$ by the product of $z^2$ and $y_k$,
and then insert the expression for $y_k$ from (\ref{yk}). After
inserting the above approximate expressions for $\beta^{\pm}_k$ we
obtain
\be
P_k \, \sim \, k^2 a^{-2}(\tau_k^0) a^{-1/2}(\tau_k^0 + l_s^2k)
a^{1/2}(\tau_k^0 - l_s^2k) \, .
\ee
It is apparent that the nonlocal coupling between background
and fluctuation mode has a large effect. Making use of
(\ref{inittime}) (and keeping in mind that the second term in
(\ref{inittime}) dominates over the first one, the final
result becomes
\be
P_k = c' k^{3 \over n+1},
\ee
where
\be
c' = \frac{ M_s^{5n-1 \over n+1} }
     { 2^{n-1 \over 2(n+1)} \alpha_0^3 M_P^2 }.
\ee
This is obviously very different from the old result (\ref{oldPk}).
Instead of a red spectrum we now obtain a blue spectrum.

To summarize, for an accelerating universe ($n > 1$),
the spectrum has a negative spectral index in the UV region,
but a positive one in the IR region.
A smooth interpolation between the two regions
would yield a nearly flat spectrum in the transition region.
Note that as $n$ increases, the slope of the power spectrum
decreases. Thus, the more rapidly the Universe is accelerating,
the closer the spectrum is to being scale-invariant. This is
similar to what occurs in ordinary power law inflation, except that
the sign of the spectral index is opposite. In the limit towards
exponential inflation, the results for commutative and non-commutative
space-times converge.

The case of exponential inflation is very special
in that we do not need to specify $z_k$ or $\tilde{z}_k$
to determine the index of the power spectrum.
Despite the fact that $z_k(\e)$ and $\tilde{z}_k(\e)$
are different from $a$,
their dependence on $k$ happens to be the same as $a(\e^0_k)$:
$z_k(\e^0_k) \propto \tilde{z}_k(\e^0_k) \propto a(\e^0_k) \propto k$.
Therefore the spectrum is scale invariant,
like in the commutative case.
For $H \ll M_s$, the approximation (\ref{app1}) is valid,
and the spectrum is roughly the same as (\ref{infl}).
But for $H \gg M_s$, the magnitude can be different.
For instance, if we take (\ref{zkdef}), it is
\be \label{infl2}
P_k \simeq \frac{ M_s^5 }{\sqrt{2} H^3 M_P^2}.
\ee
If we assume that (\ref{infl}) is correct
for the tensor metric perturbations,
the observational bound for gravitational waves
$P_k < 10^{-10}$
leads to the usual hierarchy problem \cite{Hbound}
$H < 10^{-5} M_P$
of inflationary cosmology.
But if (\ref{infl2}) is correct,
the problem is alleviated.
For $H \simeq M_P$, we only need $M_S < 10^{-2} M_P$.

\section{Discussion}

We have studied the consequences of the SSUR on the evolution
of cosmological fluctuations in expanding cosmological
backgrounds. Given a noncommutative spacetime obeying the
SSUR, the cosmological background will still be described
by the Einstein equations since the background fields only
depend on one spacetime variable. The equations for the
linear fluctuations, however, are modified. We have argued
that the modifications take the form of an interaction of
the fluctuating field with the background which is nonlocal
in time.

An important consequence of the SSUR is that for each
comoving wavenumber $k$, there is an earliest time $\e_k^0$ at
which the fluctuating mode exists. We assume that the
fluctuation starts out with its vacuum amplitude at this
time. Since the dependence of $\eta_k^0$ on $k$ is nontrivial,
we expect that the index of the power spectrum of cosmological
fluctuations will be different than for commutative spacetime.
We find that this is indeed the case
for a range of wavelengths,
except in the special case of exponential inflation
in which we also obtain a scale-invariant spectrum.
Note that for power law inflation ($n > 1$)
the spectrum of fluctuations is ``blue''
when the SSUR is effective
(for the infra-red region $k < A$)
in contrast to the case of a commutative spacetime
in which the spectrum is ``red''.
The reason for this difference is that in our case,
modes with smaller values of $k$ are generated later
than those with a larger value,
and thus experience growth due to squeezing for less time,
whereas for commutative spacetime the larger the value of $k$,
the later the mode leaves the Hubble radius and
the less squeezing it will experience. 

Let us now conclude with a few remarks.

\begin{enumerate}
\item
Although we find the same requirement $m=-1$
as in the undeformed scalar field theory
for a scale-invariant spectrum,
it represents a special case because
the spectrum is different
for other values of $m$ in the IR.
\item
In the case of exponential inflation, 
although the spectrum is always roughly scale-invariant,
the amplitudes are different for large $H$ and small $H$,
\item
It may appear a little counter-intuitive
that it is the IR fluctuation modes
that are severly modified by
the stringy spacetime uncertainty relation.
This may be interpreted as a manifestation of
the UV/IR connection.
\item
A realization of inflation
in the context of noncommutative spacetime obeying
the SSUR would not be subject to the trans-Planckian
problem of general inflationary models \cite{RHBrev}.
The case of exponential inflation in our model can also
be viewed as a test of the robustness of inflation,
but with a different kind of deformation
of the dispersion relation than those considered in
\cite{MB}.
\item
In the case of scalar metric fluctuations, we had to
make a choice of which variable to consider as the
one whose action is subject to the transformation
from commutative spacetime to noncommutative spacetime
outlined in Section III. It would be interesting to
study the results for other choices. 
\item
An intriguing possibility is that there is
only one effective scale factor, i.e., $z_k = \tilde{z}_k$.
With this assumption, we do not need
the approximation (\ref{app1}) to calculate the spectrum.
The spectrum (\ref{Pk2}) can be interpreted as
the physical energy $E_k$ squared at $\e^0_k$
for all modes with $\e^0_k > \e^1_k$.
The SSUR implies that $E_k(\e^0_k) = M_s$,
and so we obtain
\be
P_k \simeq \frac{M_s^2}{M_P^2}.
\ee
for all UV modes (\ref{b1}).

\end{enumerate}

\section*{Acknowledgments}

RB thanks Karim Benakli and Yaron Oz
for useful discussions,
and is grateful to Prof. P. Hwang
for the invitation to deliver a series of
lectures in Taiwan.
PMH thanks Tzihong Chiueh,
Je-An Gu, Hsien-chung Kao,
Feng-Li Lin, Sanjaye Ramgoolam, Hyun-Seok Yang
for helpful discussions.
The work of RB is
supported in part by the U.S. Department of Energy
under Contract
DE-FG02-91ER40688, TASK A, and by the CERN Theory
Division. RB thanks the CERN Theory Division for
hospitality during the course of this work.
The work of PMH is supported in part by
the National Science Council,
the CosPA project of the Ministry of Education,
the National Center for Theoretical Sciences,
Taiwan, R.O.C.
and the Center for Theoretical Physics
at National Taiwan University.

\section*{Appendix: Essentials of the Theory of Cosmological Perturbations}

For a detailed review of the calculation
of metric perturbations,
the reader is directed to other references \cite{MFB,LLbook}
(see also \cite{Brandenberger:2002hs} for a recent short review).
Here we outline the basic steps
with modifications due to
noncommutative spacetime or SSUR effects.

There are two kinds of metric perturbations of interest in early
Universe cosmology: the scalar and tensor fluctuations.
Tensor fluctuations correspond to gravitational waves. The perturbed
metric only has nonvanishing space-space components $h_{i j}$ which
can be expanded in terms of the two basic traceless and symmetric
polarization tensors $e^{+}_{i j}$ and $e^{x}_{i j}$ as
\be
h_{i j} \, = \, h_{+}e^{+}_{i j} + h_{x}e^{x}_{i j} \,
\ee
where the space and time dependence is in the coefficient functions
$h_{+}$ and $h_{x}$.

The Einstein action can be expanded to second order in the
metric fluctuations about a Friedmann-Robertson-Walker (FRW) 
background (\ref{FRW}),
and the action for $h_{+}$ and $h_{x}$ reduces to that of a
free, massless, minimally coupled scalar field in the FRW background.
To obtain the correct normalization, the metric must be multiplied by
the normalization factor $M_{pl}/\sqrt{2}$. In Fourier space,
the action is
\be \label{gravac}
S=\int d\eta \frac{1}{2} a^{d-1}\left(
\varphi_{-k}^{'}\varphi_k^{'} - k^2\varphi_{-k}\varphi_k\right) \, ,
\ee
where $\varphi$ stands for the coefficient functions $h_{+}$ and $h_{x}$,
and $d$ denotes the number of spatial dimensions (set to $3$ in the
following in most of this Appendix).
This leads to the equation of motion 
\be
\varphi_k^{''} + 2 {{a^{'}} \over a} \varphi^{'}_k + k^2 \varphi_k \, 
= \, 0 \, .
\ee
The Hubble friction term can be eliminated via a change of variables
\be
\mu \, = \, a \varphi \, ,
\ee
yielding the equation of motion
\be \label{graveq}
\mu_k^{''} + \bigl( k^2 - {{a^{''}} \over a} \bigr) \mu_k \, = \, 0 \, .
\ee

The power spectrum of gravitational waves in a particular state
$|0>$ of the gravitational radiation field can be written in terms
of the new field $\mu$ as
\be \label{gravpow}
{\cal P}_g(k) \, 
= \, 2 {{k^3} \over {2 \pi^2}}{{<0|\mu_k^{*}\mu_k|0>} \over {a^2}} \, .
\ee
The two point function appearing in (\ref{gravpow}) is that of a free
canonically normalized massless scalar field multiplied by $2 / M_{pl}^2$.

Scalar metric fluctuations couple to matter, and give rise to the
large-scale structure of the Universe. The description of scalar
metric perturbations is more complicated than the analysis of
gravitational waves both because of the coupling to matter and also
because some perturbation modes correspond to space-time reparametrizations
of a homogeneous and isotropic cosmology. This is the issue of gauge
fixing.
A simple way to address this issue is to work in a system of
coordinates which
completely fixes the gauge. A simple choice is the {\it longitudinal}
gauge, in which the metric takes the form \cite{MFB}
\be
ds^2 \, = \, a^2(\eta) \bigl[(1 + 2 \Phi) d\eta^2 - (1 - 2 \Psi) \gamma^{i j}
dx_i dx_j \bigr] \, ,
\ee
where the space and time dependent functions $\Phi$ and $\Psi$ are the two
physical metric degrees of freedom which describe scalar metric fluctuations
($\gamma^{i j}$ is the metric of the unperturbed spatial hypersurfaces).
The fluctuations of matter fields give additional degrees of freedom for
scalar metric fluctuations. In the simple case of a single scalar matter
field, the matter field fluctuation can be denoted by $\delta \varphi$.
In the absence of anisotropic stress, it follows from the Einstein equations
that the two metric fluctuation variables $\Phi$ and $\Psi$ coincide. Due
to the Einstein constraint equation, the remaining metric fluctuation
$\Psi$ is determined by the matter fluctuation $\delta \varphi$. 

It is
clear from this analysis of the physical degrees of freedom that the
action for scalar metric fluctuations must be expressible in terms
of the action of a single free scalar field $v$ with a time dependent mass 
(determined by the background cosmology). As shown in \cite{M85} (see
also \cite{L80}, this field is
\bea
v \, &=& \, a \bigl( \delta \varphi + {{\varphi_0^{'}} \over {\cal H}} \Psi 
\bigr) \nonumber \\
&=& z {\cal R} \, ,
\eea
where $\varphi_0$ denotes the background value of the scalar matter field, 
${\cal H} = a^{'} / a$, 
\be \label{zdef}
z \, = \, a {{\varphi^{'}_0} \over {\cal H}} \, ,
\ee
and ${\cal R}$ denotes the curvature perturbation in comoving gauge \cite{L85}.
The action for scalar metric fluctuations is \cite{M88}
\be \label{scaleac}
S \, = \, {1 \over 2} \int d^4x \sqrt{-\gamma} 
\bigl( v^{'2} - \gamma^{i j} v_{, i} v_{, j} + {z^{''} \over z} v^2 \bigr) \, ,
\ee
(where $\gamma$ is the determinant of the metric $\gamma^{i j}$)
which leads to the equation of motion
\be
v^{''}_k + \bigl( k^2 - {{z^{''}} \over z} \bigr) v_k \, = \, 0 \, ,
\ee
which under the change $a \rightarrow z$ is identical to the equation
(\ref{graveq}) for gravitational waves. Note that if $a(\eta)$ is a power
of $\eta$, then $\varphi^{'}_0$ and ${\cal H}$ scale with the same
power of $\eta$ so that $z$ is proportional to $a$, in which case
the evolution of gravitational waves and scalar metric fluctuations is
identical.

In analogy to (\ref{gravpow}), the power spectrum of the curvature
fluctuation ${\cal R}$ is
\be
{\cal P}_{\cal R}(k) \, 
= \, {{k^3} \over {2 \pi^2}}{{<0|v_k^{*}v_k|0>} \over {z^2}} \, .
\ee

So far, the theory was developed for commuting space-time variables.
To make the transition to non-commutative space-time, and to take
into account the space-time uncertainty relation, we replace, as
discussed in the main part of the text, the $k-$independent function
$z(\eta)$ by a $k-$dependent function $z_k(\eta)$. In addition, we
allow the initial time $\eta^0$ to depend on $k$.

In order to show how the growth of the classical mode function $v_k$
translates into the growth of the expectation value which determines
the power spectrum we will make use of the Hamiltonian formalism.
From the action (\ref{scaleac}) it follows that the momentum canonically
conjugate to the field $v$ is
\be
\Pi_k \, = \, v^{'}_{-k} - {{z_k^{'}} \over {z_k}} v_{-k} \, ,
\ee
and this leads to the Hamiltonian (see e.g.\cite{MB})
\be \label{H}
H \, = \, \int d^d k\left[k a_k^{\dagger}a_k
+\frac{i}{2}\frac{z^{'}_k}{z_k}
(a_k^{\dagger}a_{-k}^{\dagger}-a_k a_{-k})
\right],
\ee
where $a_k^{\dagger}(\eta)$ and $a_k(\eta)$
are the creation and annihilation operators at time $\eta$
related to $v_k$ and its conjugate momentum by
\bea \label{creation}
v_k \, &=& \, \frac{1}{\sqrt{2k}}(a_k + a_{-k}^{\dagger}), \\
\Pi_k \, &=& \, \frac{-i}{\sqrt{2k}}(a_{-k}-a_k^{\dagger}).
\eea

In the Heisenberg picture,
the creation and annihilation operators evolve with time,
while the state does not.
Assume that the Universe is in
the state $|0>$ defined by
\be
a_k(\eta^0_k)|0> \, = \, 0 \, ,
\ee
which is the vacuum for mode $k$ at some intial time $\eta^0_k$.
It is in general not the vacuum at later times.

The Bogoliubov transformation relates the creation and
annihilation operators at $\eta^0$ with the corresponding operators
at the time $\eta$:
\bea
a_k(\eta) &=&
\alpha_k(\eta)a_k(\eta_0)+\beta_k(\eta)a_{-k}^{\dagger}(\eta_0), \\
a_k^{\dagger}(\eta) &=& 
\bar{\beta}_k(\eta)a_{-k}(\eta_0)+\bar{\alpha}_k(\eta)a_k^{\dagger}(\eta_0) 
\, ,
\eea
where $\alpha_k$ and $\beta_k$ satisfy
\be \label{uv}
\alpha_k\bar{\alpha}_k - \beta_k\bar{\beta}_k=1, \quad \forall k \, .
\ee
The Hamilton equations
$[H, a_k]=i\dot{a}_k$ and its Hermitian conjugate take on a
simpler form when written in terms of the new variables
\be \label{newvar}
\zeta_k = \alpha_k-\bar{\beta}_k, \quad \xi_k = \alpha_k+\bar{\beta}_k.
\ee
In terms of them they read
\bea
&\zeta^{''}_k + 
\left(k^2-\frac{z^{''}_k}{z_k}\right)\zeta_k=0,
\label{eq1} \\
&\xi^{''}_k + 
\left(k^2-\frac{(z^{-1}_k)^{''}}{(z^{-1}_k)}\right)\xi_k=0,
\label{eq2}
\eea
The larger the value of $|\beta_k|^2(\eta)$, the larger the number
of particles at time $\eta$ created out of the initial vacuum state $|0>$.

Let us denote the factors in (\ref{eq1}), (\ref{eq2}) by
\be
M=k^2-\frac{z^{''}_k}{z_k}, \quad
N=k^2-\frac{(z^{-1}_k)^{''}}{(z^{-1}_k)}.
\ee
For small $\eta$ ($\eta<-1/k$) 
(scales smaller than the Hubble radius),
both $M$ and $N$ are positive and approximately equal to $k^2$.
Thus, $\zeta_k$ and $\xi_k$ oscillate.
If the initial state is taken to be the local vacuum state,
\be
\alpha_k(\eta_0)=1, \quad \beta_k(\eta_0)=0,
\ee
then the magnitudes of $\zeta$ and $\xi$ are of order $1$
until $\eta=-1/k$.
This represents the oscillation of quantum vacuum fluctuations.
We will refer to this regime as phase $I$.
On scales larger than the Hubble radius (at later times), 
$M$ and $N$ are dominated by the second (negative) term.
In this period, $\zeta$ and $\xi$ correspond to frozen
fluctuations which are undergoing quantum squeezing and
which scale like $z_k\sim a$ and $z^{-1}_k\sim a^{-1}$,
respectively.
Since $a$ is increasing with time,
by (\ref{newvar}) we can approximate $v_k$ by $\zeta_k/\sqrt{2k}$
for sufficiently late times.
This is phase $\II$
When $k^2$ is much smaller than both
$z^{''}_k/z_k$ and $(z^{-1}_k)^{''}/(z_k^{-1})$,
and assuming that $z$ is an increasing function,
the dominant solutions are simply
\be
\zeta_k=C_k z_k, \quad \xi_k=\frac{1}{C_k z_k}
\ee
for real $u_k$, $v_k$.
Note that the condition (\ref{uv})
is satisfied for this solution,
but not for other solutions of
the second order differential equations.
Suppose that for sufficiently late times,
$z_k\gg 1$ ($z_k\ll 1$)
then $v_k\simeq \frac{1}{2}\zeta_k/\sqrt{2k}$
($v_k\simeq \frac{1}{2}\xi_k/\sqrt{2k}$).
Since the initial condition is
$\zeta_k(\eta_0)=\xi_k(\eta_0)=1$,
it is equivalent to say that
for $z_k \gg 1$,
for sufficiently large $\eta$,
$v_k(\eta)$ is given by
the solution to the differential equation (\ref{eq1})
with the initial condition $v(\eta_0)=1/\sqrt{2k}$.
For $z_k \ll 1$,
we replace $z_k$ by $z^{-1}_k$.
In general we want to find the larger of the two
functions $\zeta_k$ and $\xi_k$.
This then determines via the relations(\ref{newvar})
the Bogoliubov coefficient $\beta_k$,
in terms of which the power spectrum of metric fluctuations becomes
\be \label{spect}
{\cal P}_{\cal R}(k) \, = \,
{{k^3} \over {2 \pi^2}}{1 \over {2k}} {{|\beta_k|^2} \over {z_k^2}} \, .
\ee
Note that the factor $1 / (2k)$ represents the vacuum normalization
of the states, seen in (\ref{creation}).

In summary, for expanding $z_k$,
the spectrum is
\be \label{Pk}
{\cal P}_{\cal R}(k) \, = \,
{k^2 \over 4 \pi^2} {1 \over z_k^2(\eta_k)},
\ee
where $\eta_k$ is the time when
the fluctuation mode $k$ crosses the Hubble radius ($M = 0$).
In this paper we also consider the case of
fluctuations outside the Hubble radius starting in the vacuum.
For them $\eta_k$ should be taken to be the time when they
start in the vacuum.




\end{document}